\documentclass[12pt,preprint]{aastex}

\usepackage{natbib}
\usepackage{color}
\usepackage{comment}

\usepackage{bm}
\usepackage{lscape}

\newcommand{\ergs}{erg~s$^{-1}$ }

\newcommand{\Co}{$^{56}$Co }
\newcommand{\Ni}{$^{56}$Ni }

\newcommand{\lam}{$\lambda$ }

\shorttitle{Luminous Type Ib SN 2012au}
\shortauthors{Takaki et al.}

\begin{document}

\title{
A Luminous and Fast-Expanding Type Ib Supernova SN 2012au
}

\author{
Katsutoshi Takaki\altaffilmark{1},
Koji S. Kawabata\altaffilmark{2},
Masayuki Yamanaka\altaffilmark{3},
Keiichi Maeda\altaffilmark{4},
Masaomi Tanaka\altaffilmark{5},
Hiroshi Akitaya\altaffilmark{2},
Yasushi Fukazawa \altaffilmark{1,2},
Ryosuke Itoh\altaffilmark{1},
Kenzo Kinugasa\altaffilmark{6},
Yuki Moritani\altaffilmark{2},
Takashi Ohsugi\altaffilmark{2},
Mahito Sasada\altaffilmark{7},
Makoto Uemura\altaffilmark{2},
Issei Ueno\altaffilmark{1},
Takahiro Ui\altaffilmark{1},
Takeshi Urano\altaffilmark{1},
Michitoshi Yoshida\altaffilmark{2},
Ken'ichi Nomoto\altaffilmark{4}
}

\altaffiltext{1}{Department of Physical Science, Hiroshima University, 1-3-1 Kagamiyama, Higashi-Hiroshima,
Hiroshima 739-8526, Japan; takaki@hep01.hepl.hiroshima-u.ac.jp}
\altaffiltext{2}{Hiroshima Astrophysical Science Center, Hiroshima University, 1-3-1 Kagamiyama,
Higashi-Hiroshima, Hiroshima 739-8526, Japan}
\altaffiltext{3}{Kwasan Observatory, Kyoto University, Ohmine-cho Kita Kazan, Yamashina-ku, Kyoto 607-8471, Japan}
\altaffiltext{4}{Kavli Institute for the Physics and Mathematics of the Universe (WPI),
Todai Institutes for Advanced Study, The University of Tokyo, 5-1-5 Kashiwanoha, Kashiwa, Chiba 277-8583, Japan}
\altaffiltext{5}{National Astronomical Observatory of Japan, 2-21-1 Osawa, Mitaka, Tokyo 181-8588, Japan}
\altaffiltext{6}{Nobeyama Radio Observatory, National Astronomical Observatory of Japan, 462-2 Nobeyama, Minamimaki, Nagano 384-1305, Japan}
\altaffiltext{7}{Department of Astronomy, Kyoto University, Kitashirakawa-Oiwake-cho, Sakyo-ku, Kyoto 606-8502, Japan}

\begin{abstract}
We present a set of photometric and spectroscopic observations of a bright Type Ib supernova SN 2012au from $-6$ d until $\sim +150$ d after maximum.
The shape of its early $R$-band light curve is similar to that of an average Type Ib/c supernova.
The peak absolute magnitude is $M_{\rm R}=-18.7 \pm 0.2$ mag, which suggests that this supernova belongs to a very luminous group among Type Ib supernovae.
The line velocity of \ion{He}{1} $\lambda$5876 is about $15,000$ km s$^{-1}$ around maximum, which is much faster than that in a typical Type Ib supernova.
From the quasi-bolometric peak luminosity of $(6.7\pm 1.3) \times 10^{42}$ \ergs, we estimate the \Ni mass produced during the explosion as $\sim$0.30 M$_{\odot}$.
We also give a rough constraint to the ejecta mass $5$--$7$ M$_{\odot}$ and the kinetic energy $(7$--$18) \times 10^{51}$ erg.
We find a weak correlation between the peak absolute magnitude and \ion{He}{1} velocity among Type Ib SNe.
The similarities to SN 1998bw in the density structure inferred from the light curve model as well as the large peak bolometric luminosity suggest that SN 2012au had properties similar to energetic Type Ic supernovae.
\end{abstract}

\section{Introduction}
Type Ib supernovae (SNe Ib) are one of subtypes of core-collapse supernovae (CC-SNe), whose early spectra are characterized by strong helium features around maximum \citep[see][for review]{1997ARA&A..35..309F}.
SNe Ib belong to a larger group of `stripped-envelope' CC-SNe \citep{1997ApJ...491..375C}, together with SNe Ic (SNe without hydrogen nor helium) and SNe IIb (showing hydrogen only during early phase).
It is commonly accepted that a progenitor star of a stripped-envelope CC-SN has lost its hydrogen envelope via either a strong stellar wind in a Wolf-Rayet (WR) phase or an interaction with a binary companion.
Deep pre-imaging studies suggest that the former, massive WR path is not always the case \citep[e.g.,][]{2005MNRAS.360..288M,2008MNRAS.391L...5C}.
Also, all SNe associated with gamma-ray burst (GRBs) have been found to be SNe Ic, with the most solid cases belonging to broad-line SNe Ic (SNe Ic-BL), attributed to a large kinetic energy of the expansion (sometimes called hypernovae).
But details of the connection are still under debate.

In recent years, many detailed observations have been performed for SNe Ib, and a large diversity has been recognized, e.g., 
a normal SN Ib SN 1990I \citep{2004A&A...426..963E},
a mildly energetic SN 2008D \citep{2008Sci...321.1185M,2009ApJ...692.1131T,2009ApJ...702..226M}, 
transitional SNe Ib/IIb SN 2008ax \citep{2011MNRAS.413.2140T}, 1999dn \citep{2011MNRAS.411.2726B}, and 2011ei \citep{2013ApJ...767...71M}, 
a rapidly-evolving faint SN 2007Y \citep{2009ApJ...696..713S}, 
faint Ca-rich SNe 2005E \citep{2010Natur.465..322P} and 2005cz \citep{2010Natur.465..326K}, 
a slowly-evolving SN 2009jf \citep{2011ASInC...3..126S}, 
an intermediate sample between SNe Ib and Ic SN 1999ex \citep{2002AJ....124..417H}, 
and a very peculiar SN 2005bf \citep{2005ApJ...631L.125A,2005ApJ...633L..97T,2007ApJ...666.1069M}.
In this Letter, we report optical photometry and spectroscopy of SN Ib 2012au in the early phase and discuss the results based on the observations.
We show that SN 2012au has observational properties similar to GRB-associated energetic SNe Ic rather than normal SNe Ib/c.

\section{Observation and Data Reduction}
SN 2012au was discovered by Catalina Real-Time Transient Survey SNHunt project on 2012 Mar 14 UT \citep{2012CBET.3052....1H} in NGC 4790 \citep[$d=23.6 \pm 0.5$ Mpc;][]{1988Sci...242..310T,2007A&A...465...71T}, and subsequently classified as SN Ib \citep{2012CBET.3052....2S,2012ATel.3968....1S,2013arXiv1304.0095M}.
We carried out photometric (54 nights) and spectroscopic (19 nights) observations of SN 2012au from 2012 Mar 15 through Aug 19 with HOWPol \citep{2008SPIE.7014E.151K} attached to 1.5m Kanata telescope at Higashi-Hiroshima Observatory.
We used $B,\ V,\ Rc,\ Ic$ and $z'$ filters for the photometric observation.
We performed PSF-photometry in each obtained image.
Landolt standard fields were used for photometric calibration of several nearby comparison stars.
For spectroscopy, we calibarated flux scale using spectrophotometric standard star HR 4963 obtained in the same nights.
We used the sky emission lines simultaneously recorded on the SN spectra for wavelength calibration, and then achieved a wavelength error of $\sim 3.5$\AA\ over wavelength range 4500--9200\AA\ with the resolution of $R\sim 400$.
It is noted that the bright nucleus of the host galaxy exists only $5''$ east from this SN and it may contaminate our photometry in the latest phase.
However, this effect would be negligible in our discussion.
In addition to the spectra of SN 2012au, we present our unpublished spectra of SN 2009jf (two epochs) obtained with GLOWS installed to the 1.5m telescope at Gunma Astronomical Observatory.

\section{Results}
\subsection{Extinction and Light Curves}
We first correct Galactic extinction of $E(B-V)_{\rm MW} = 0.048$ mag \citep{1998ApJ...500..525S}.
For the extinction within the host galaxy, we place a limit as $E(B-V)_{\rm host}\leq 0.035$ mag from the upper-limit of the equivalent width of \ion{Na}{1} D absorption line \citep{2012MNRAS.426.1465P} in the averaged spectra. 
The extinctions in the host galaxy is thus negligibly small ($\lesssim 0.1$ mag in $V$ band)
\footnote{This is consistent with $E(B-V)_{\rm host} = 0.02 \pm 0.01$ mag suggested by \cite{2013arXiv1304.0095M}.}.
Therefore, we adopt the total extinction toward SN 2012au as $E(B-V)_{\rm total} = 0.048$ mag.

The light curves (LCs) and color curves are shown in Figure 1.
The SN reached peak brightness $R_{\rm max}=13.1\pm 0.1$ mag on Mar $21.0 \pm 1.0$ (which is set to be $t=0$ d), corresponding to the absolute magnitude $M_{R,\ {\rm max}} = -18.7 \pm 0.2$ mag.
A compilation by Lick Observatory Supernova Search (LOSS) indicates that the mean absolute magnitude of SNe Ib/c is $\langle M_{R,\ {\rm max}}\rangle = -16.1 \pm 1.2$ mag \citep{2011MNRAS.412.1441L}.
Another systematic study suggests that $\langle M_{R,\ {\rm max}}\rangle = -17.9 \pm 0.9$ mag for SNe Ib \citep[and $-18.3\pm 0.6$ mag and $-19.0\pm 1.1$ mag for SNe Ic and SNe Ic-BL, respectively;][]{2011ApJ...741...97D}.
SN 2012au belongs to a very luminous group among SNe Ib.
The post maximum decline within 15 d is estimated to be $\Delta m_{15}(R) = 0.57 \pm 0.06$ mag, which locates near the center of the cluster of Drout et al.'s samples, which ranges from $\sim$0.4 to $\sim$0.8 mag.
After $t\sim 30$ d, the SN showed a slow decline with the rate $0.017$ mag d$^{-1}$ (average in 34--111 d) in $R$ band.

\begin{figure*}
\begin{center}
\includegraphics[scale=0.80,angle=-90]{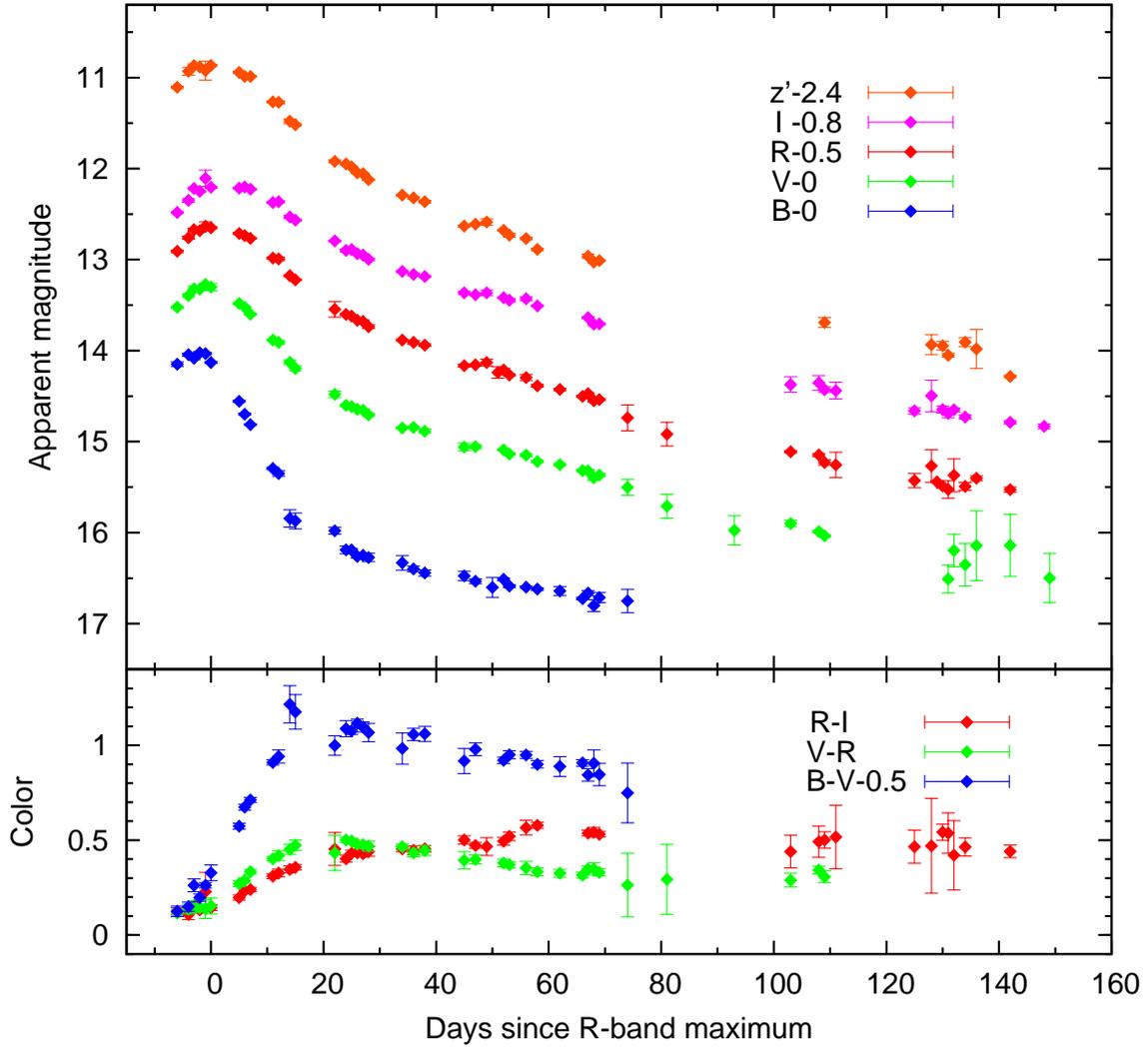}
\caption{Top panel: $BVRIz'$ light curves of SN 2012au with magnitude offsets to avoid overlap, as indicated in the panel. The extinction of $E(B-V)_{\rm total}=0.048$ mag has been corrected for. Bottom panel: Color evolutions in $B-V$, $V-R$, and $R-I$.}
\end{center}
\end{figure*}

In the upper panel of Figure 2, we show a comparison of $R$-band LC with other CC-SNe, SN 2008ax (Ib/IIb), SN 2009jf (Ib), SN 1998bw (Ic-BL) and SN 1993J (IIb) and also with the LOSS averages.
Around maximum, the LC of SN 2012au is very similar to those of SN 1998bw and the LOSS average, showing a slightly slower evolution than LCs of SNe 2008ax and SN 1993J.
On the other hand, in the tail of LC ($t\gtrsim 30$ d), SN 2012au shows slower evolution than these SNe except for the slowly-evolving SN Ib 2009jf.
These facts suggest that the trapping efficiencies of optical and $\gamma$-ray photons within the ejecta are larger than the typical.

\begin{figure*}
\begin{center}
\includegraphics[scale=0.80,angle=-90]{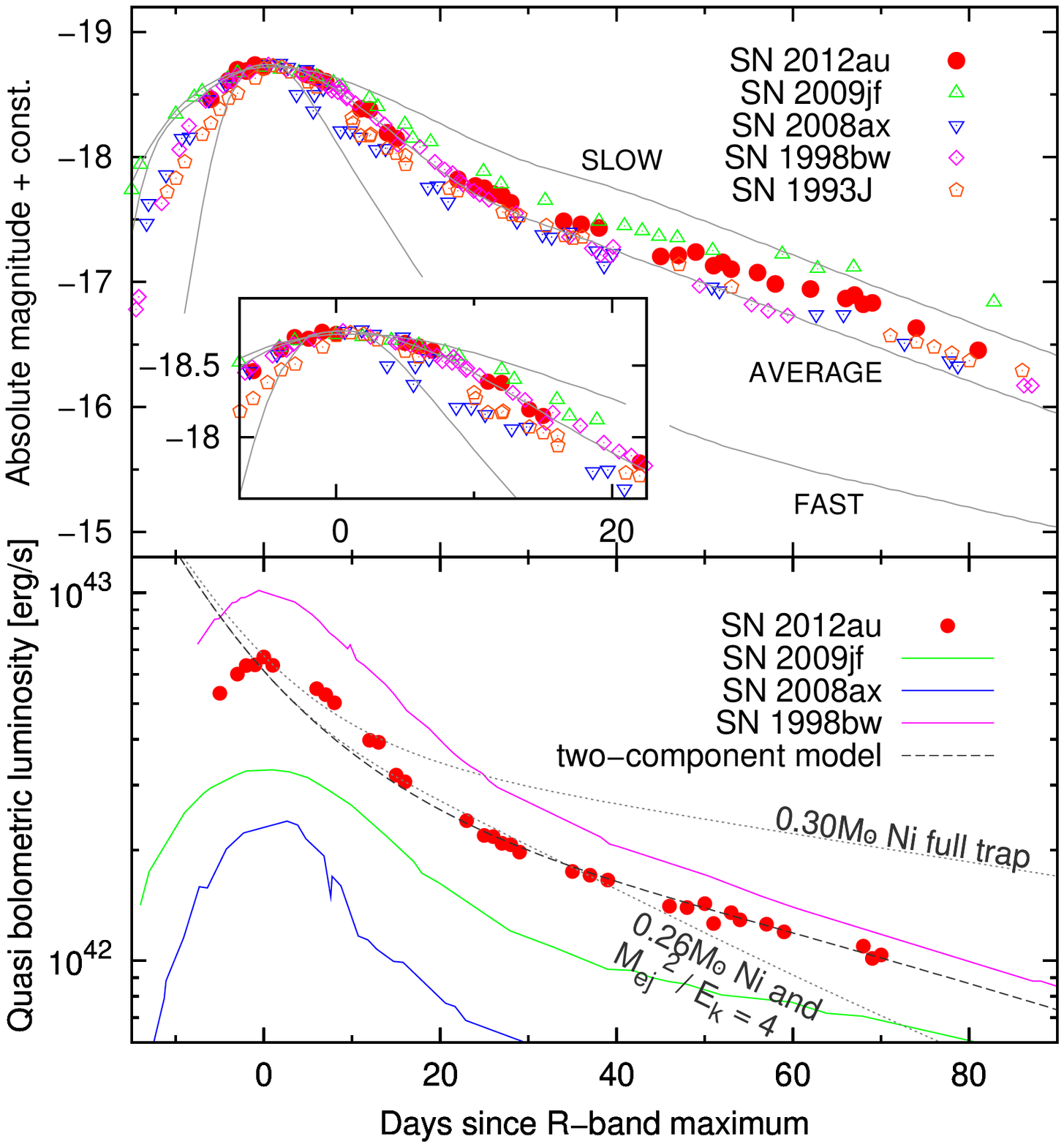}
\caption{Top panel: $R$-band light curve of SN 2012au comparared with those of SN 2009jf \citep[Ib;][]{2011ASInC...3..126S}, SN 2008ax \citep[Ib/IIb;][]{2011MNRAS.413.2140T,2009PZ.....29....2T}, SN 1998bw \citep[Ic-BL;][]{2001ApJ...555..900P} and SN 1993J \citep[IIb;][]{1996AJ....112..732R}. The lines are for slow, average, and fast LC evolutions of SN Ib/c in the LOSS samples \citep{2011MNRAS.412.1441L}. They are shifted to match to the peak of SN 2012au. Bottom panel: Quasi-bolometric LC of SN 2012au compared with those of SN 2009jf, SN 2008ax and SN 1998bw. We also show some analytic LC models (see \S 3.3).}
\end{center}
\end{figure*}

The color evolution is shown in the lower panel of Figure 1.
It becomes progressively redder until $\sim$20 d and then its slopes becomes rather flat, which is similar to those seen in other SNe Ib/c.
\cite{2011ApJ...741...97D} suggested that the intrinsic $V-R$ colors fall in a relatively narrow range $0.26 \pm 0.06$ mag around 10 d.
In case of SN 2012au, $V-R=0.36\pm 0.02$ mag at 10 d, which is slightly redder than the mean value.

\subsection{Spectra}
We show the spectral evolution from $-6$ d to $+138$ d in Figure 3.
Around maximum, the absorption line of \ion{He}{1} \lam 5876 is conspicuous as in other SNe Ib.
We can see other \ion{He}{1} lines ($\lambda\lambda$ 6678, 7065), \ion{Fe}{2} \lam 5169 and \ion{Ca}{2} IR triplet.
After $+110$ d, nebular emission lines, [\ion{O}{1}] $\lambda\lambda$ 6300,6364 and [\ion{Ca}{2}] $\lambda\lambda$ 7291,7323, appear.
A comparison of the spectra around maximum and $+35$ d are shown in the bottom panels of Figure 3.
It is clear that the blueshift of \ion{He}{1} and other lines in SN 2012au is larger than other SNe, which is discussed in \S 4.2.

\begin{figure*}
\begin{center}
\includegraphics[scale=0.85,angle=-90]{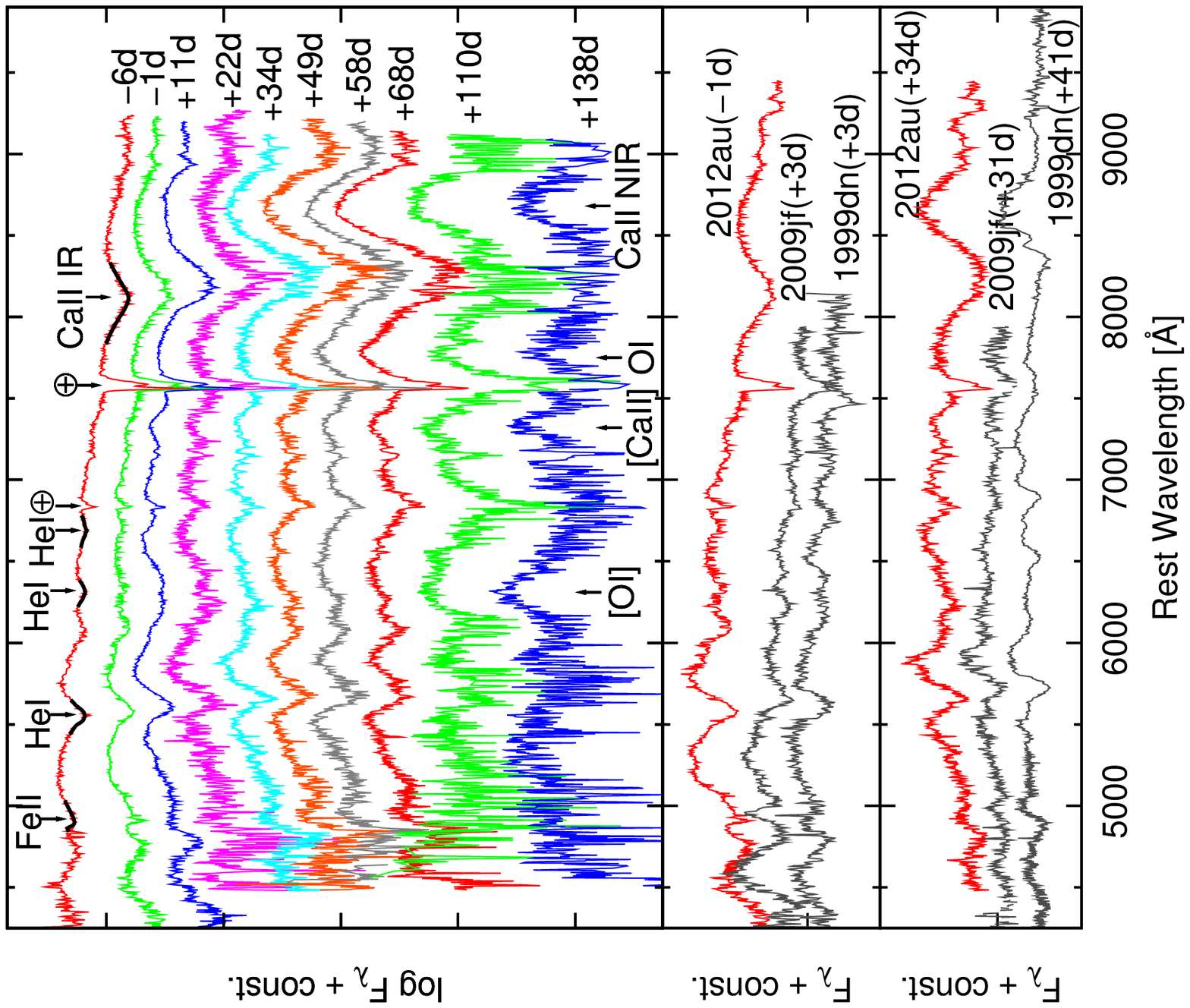}
\caption{Top panel: Spectral evolution of SN 2012au from $-6$ d to $+138$ d. Each spectrum is plotted in a logarithmic flux scale and arbitrarily shifted to avoid overlap. The interstellar extinction (\S 3.1) has not been corrected for. Identifications for several features (highlighted by thick short lines on the first spectrum) are shown, \ion{Fe}{2} $\lambda$5169, \ion {He}{1} $\lambda$5876, \ion{He}{1} $\lambda$6678, \ion{He}{1} $\lambda$7065, and \ion{Ca}{2} IR triplet. Bottom two panels: Comparison of spectra around maximum and $+35$ d with SN 2009jf (slowly-evolving Ib; our unpublished data) and SN 1999dn \citep[typical Ib;][]{2001AJ....121.1648M}, shown in linear flux scale. The reference spectra of SN 1999dn has been downloaded from SUSPECT\protect\footnotemark .}
\end{center}
\end{figure*}
\footnotetext{http://bruford.nhn.ou.edu/$\sim$suspect/}

\section{Discussion}
\subsection{Bolometric luminosity and $^{56}$Ni mass}
From our $BVRI$ photometry, we calculated quasi-bolometric luminosity, assuming that the $BVRI$ bands occupy about 60\% of the bolometric one around the peak \citep{2006ApJ...644..400T}.
Because of this simple assumption, this bolometric luminosity may well have a large systematic error ($\sim 20$\%).
The derived quasi-bolometric LC is shown in the bottom panel of Figure 2.
It reaches $(6.7\pm 1.3)\times 10^{42}$ erg s$^{-1}$ at maximum.
This is clearly larger than those of SN 2009jf \citep[$3.2 \times 10^{42}$ erg s$^{-1}$;][]{2011ASInC...3..126S} and SN 2008ax \citep[$2.4 \times 10^{42}$ erg s$^{-1}$;][]{2011MNRAS.413.2140T} \citep[cf. $1.0 \times 10^{43}$ erg s$^{-1}$ for luminous SN 1998bw;][see below]{2001ApJ...550..991N}.

A general interpretation of the radiation source of SNe Ib/c around maximum is that the energy generated by the decays of \Ni and \Co emerges out of the photosphere as optical radiation \citep{1982ApJ...253..785A}.
Therefore, the \Ni mass, $M(^{56}{\rm Ni})$, can be derived from the peak bolometric luminosity $L_{\rm max}$ and the rising time $t_r$ \citep{2006A&A...460..793S} as
\begin{equation}
 L_{\rm max} = \Big( 6.45\times e^{-\frac{t_r}{8.8}} + 1.45\times e^{-\frac{t_r}{111.3}} \Big) \times \Big( \frac{M(^{56}{\rm Ni})}{M_{\odot}}\Big) \times 10^{43} {\rm erg~s}^{-1} {\rm .}
\end{equation}
With $t_r = 16.5 \pm 1.0$ d \citep{2013arXiv1304.0095M} and $L_{\rm max}=6.7\times 10^{42}$ erg s$^{-1}$, we obtain $M(^{56}{\rm Ni}) = 0.30$ M$_{\odot}$.
This \Ni mass is larger than  that in other SNe Ib \citep[e.g., 0.07--0.15 M$_{\odot}$ in SN 2008ax and 0.14--0.20 M$_{\odot}$ in SN 2009jf; see also][]{2011ApJ...741...97D}.

Next, we try to constrain the \Ni mass from the tail component of the quasi-bolometric LC ($t=30$--$70$ d) using a simple one-zone model \citep{2003ApJ...593..931M} as shown in Figure 2.
At first, with a simplified assumption that $\gamma$-rays are fully trapped and deposit their energy in the ejecta, this model gives \Ni masses of 0.30 M$_{\odot}$ for the peak bolometric luminosity (as is equivalent to equation 1).
For a more realistic treatment where the $\gamma$-rays can only be partly absorbed, we need to provide the optical depth to $\gamma$-ray within the ejecta, $\tau$.
In this case the bolometric luminosity in the later phase ($t \gtrsim 30$ d) can be written as
\footnote{We show only \Co decay contribution here for simplicity. The model curves plotted in Figure~2 contain both \Ni and \Co decay contributions.}
\begin{equation}
 L = M(^{56}{\rm Ni}) e^{(-t_d/113 \scriptsize{\mbox{d}})} \big[ \epsilon_{\gamma}(1-e^{-\tau})+\epsilon_{e^{+}} \big]
\end{equation}
and
\begin{equation}
 \tau = 1000 \times \frac{(M_{\rm ej}/M_{\odot})^2}{E_{51}} ~t_{d}^{-2} \mbox{ ,}
\end{equation}
where $t_d$ is time after the explosion in day ($\equiv t_r + t$), $\epsilon_{\gamma} = 6.8\times 10^9 {\rm ~erg ~s^{-1} ~g^{-1}}$ and $\epsilon_{e^{+}} = 2.4\times 10^8 {\rm ~erg ~s^{-1} ~g^{-1}}$ are energy deposits by $\gamma$-ray and positrons, respectively, and $(M_{\rm ej}/M_{\odot})$ is the ejecta mass in solar mass unit.
By changing these parameters, however, it turns out that we are unable to reproduce the peak and tail simultaneously with a single component (for example, the case of $M(^{56}{\rm Ni})$ = 0.26 M$_{\odot}$ and $[(M_{\rm ej}/M_{\odot})^2/E_{51}] = 4$ is shown in figure 2).
The tail luminosity requires a small value of $\tau$, while tail slope requires a large value of $\tau$ instead.

Therefore, we introduce a two-component model similar to \citet{2003ApJ...593..931M}, consisting of an inner region having large optical thickness $\tau_{\rm in}$ and an outer region having small $\tau_{\rm out}$.
The luminosity is then expressed as
\begin{eqnarray}
 L_{\rm opt} = M_{\rm in}(^{56}{\rm Ni}) e^{(-t_d/113 \rm {d})} \big[ \epsilon_{\gamma}(1-e^{-\tau_{\rm in}}) + \epsilon_{e^{+}} \big] + M_{\rm out}(^{56}{\rm Ni}) e^{(-t_d/113 \rm{d})} \big[  \epsilon_{\gamma}(1-e^{-\tau_{\rm out}})+\epsilon_{e^{+}} \big] \mbox{,}
\end{eqnarray}
\begin{eqnarray}
 \tau_{\rm in} &=& 1000 \times \Big[ \frac{(M_{\rm ej}/M_{\odot})^2}{E_{51}} \Big]_{\rm in} ~t_{d}^{-2}  \mbox{, and}\\
 \tau_{\rm out} &=& 1000 \times \Big[ \frac{(M_{\rm ej}/M_{\odot})^2}{E_{51}} \Big]_{\rm out} ~t_{d}^{-2} \mbox{.}
\end{eqnarray}
The observation is reproduced reasonably well using this model with 
$M_{\rm in}(^{56}{\rm Ni) = 0.14~M_{\odot}}$, \\
$[(M_{\rm ej}/M_{\odot})^2/E_{51}]_{\rm in} = 20$, 
$M_{\rm out}(^{56}{\rm Ni) = 0.12~M_{\odot}}$, 
and $[(M_{\rm ej}/M_{\odot})^2/E_{51}]_{\rm out} = 2$ (the bottome panel of Figure 2).
This result favors that a dense core exists in their inner region.
It is interesting that a similar density contrast has also been derived for the luminous SN Ic-BL 1998bw which possibly produced a jet-like, asymmetric ejection \citep{2003ApJ...593..931M};
$M_{\rm in}(^{56}{\rm Ni) = 0.11~M_{\odot}}$,
$[(M_{\rm ej}/M_{\odot})^2/E_{51}]_{\rm in} = 26$,
$M_{\rm out}(^{56}{\rm Ni) = 0.44~M_{\odot}}$,
and $[(M_{\rm ej}/M_{\odot})^2/E_{51}]_{\rm out} = 1$.
The fraction of \Ni in the high velocity component is smaller in SN 2012au, which might suggest either a difference in details of the explosion or a different viewing direction \citep[see also][]{2013arXiv1304.0095M}.

In this two-component model, the sum of the \Ni mass is 0.26 M$_{\odot}$.
The difference from 0.30 M$_{\odot}$ derived from the peak luminosity suggests that the ratio of peak bolometric to radioactive luminosities, $\alpha$, is not unity, but $\alpha\simeq 1.2$ \citep[e.g.,][]{2006Natur.443..308H}.
Therefore, the \Ni mass of 0.26 M$_{\odot}$ is likely a better estimation. However, to provide a better comparison to other SNe (where $\alpha \sim$ 1 is typically assumed), we adopt 0.30 M$_{\odot}$ throughout this Letter.

\subsection{Line Velocity and Explosion Parameters}
We derived the line velocities of \ion{He}{1} \lam 5876, \ion{Ca}{2} IR triplet (assumed to the absorption peak at 8571 \AA\ ) and \ion{Fe}{2} \lam 5169 by fitting a quadratic-function to each absorption feature (Figure 4).
Around maximum, the He velocity of SN 2012au is $\sim 15,000$ km s$^{-1}$, which is significantly larger than those of the SNe Ib samples in \citet{2002ApJ...566.1005B} that clustered around $\sim 11,000$ km s$^{-1}$.
After 30 d, the He velocities in most SNe Ib decrease and tend to converge at $7,000$--$8,000$ km s$^{-1}$ \citep[Fig. 4; see also][]{2002ApJ...566.1005B}.
On the other hand, SN 2012au exhibits large He velocity, $\sim 10,000$ km s$^{-1}$ even at $+50$ d.
We also derived the line velocities at \ion{He}{1} $\lambda$6678 and $\lambda$7065 up to $+10$ d and confirmed that their velocities are consistent with that of \ion{He}{1} $\lambda$5876 within $1,000$ km s$^{-1}$, suggesting that the contamination by \ion{Na}{1} D to \ion{He}{1} $\lambda$5876 is negligible during the period.
The line velocities of \ion{Ca}{2} IR triplet and \ion{Fe}{2} \lam 5169 are also as large as that of \ion{He}{1} at $-6$ d (Fig. 4).
The Fe velocity shows a steep decrease from $12,500$ km s$^{-1}$ around maximum to an asymptotic value of $\sim 4,000$ km s$^{-1}$ after $+50$ d.
This trend is also seen in other SNe Ib, while the velocity in other SNe Ib is lower than $9,000$ km s$^{-1}$ at maximum \citep{2002ApJ...566.1005B}.
This may suggest that the distribution of iron is rather widespread, up to the outermost region, in the ejecta.

We show the equivalent width (EW) of \ion{He}{1} $\lambda$5876 absorption line in Figure 4.
It is about 80 \AA\ at $-6$ d, which is larger than that of SN 2008D at similar phase.
This might be related with the He enevelope mass and/or the distribution of \Ni.

\begin{figure*}
\begin{center}
\includegraphics[scale=0.62,angle=-90]{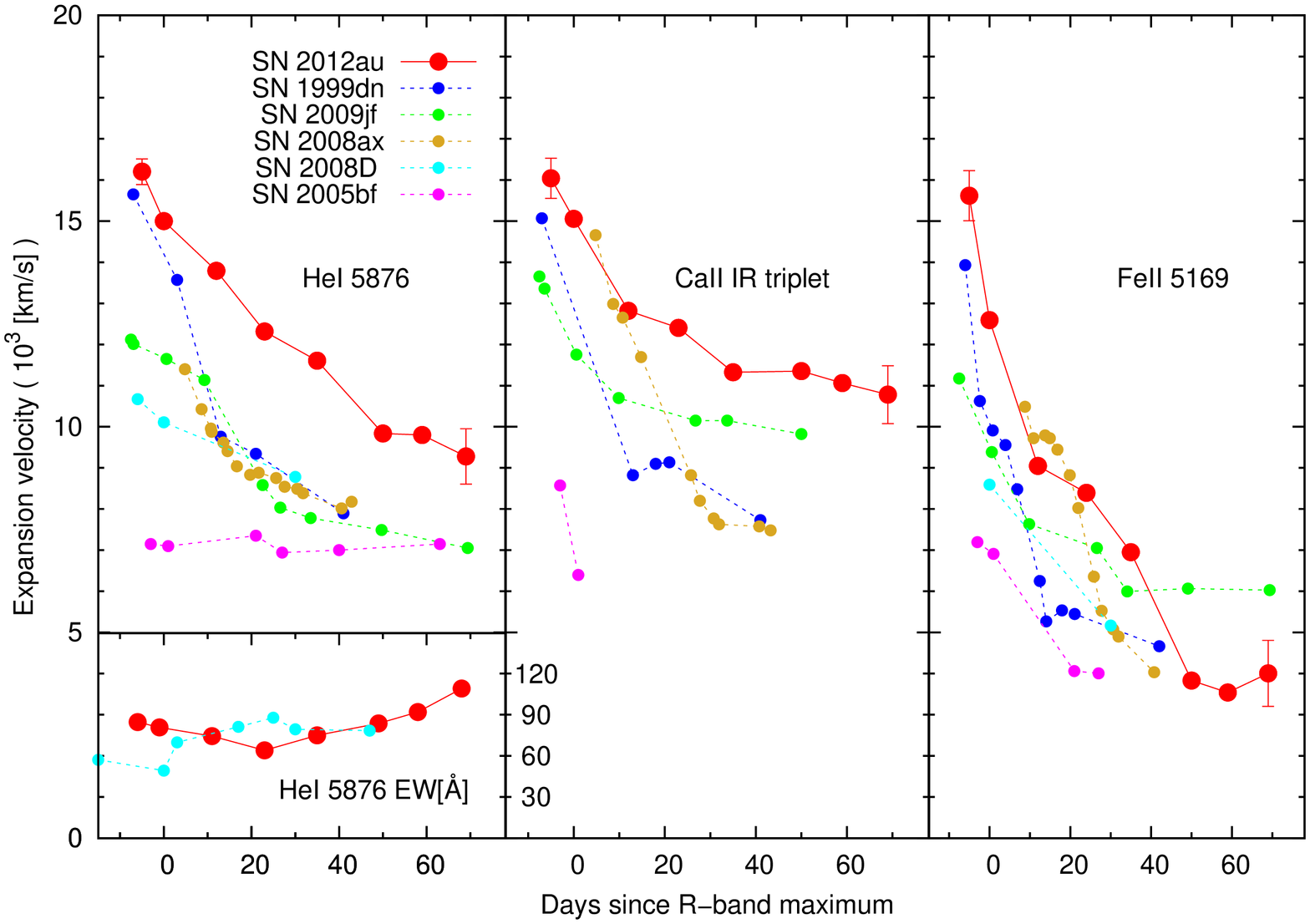}
\caption{Line velocities of \ion{He}{1} $\lambda$5876, \ion{Ca}{2} IR $\lambda$8571, and \ion{Fe}{2} $\lambda$5169 in SN 2012au compared with those of other SNe Ib. A typical error in each data point is 500--$1,000$ km s$^{-1}$. The references are common with figure 5, except for SN 1999dn \citep{2000ApJ...540..452D}, SN 1999ex \citep{2002AJ....124..417H}, SN 2008D \citep{2009ApJ...702..226M}. The error bars in the first and last data points denote typical uncertainties (1-$\sigma$) of our velocity estimation.}
\end{center}
\end{figure*}

We estimate the ejetcta mass, $M_{\rm ej}$, and kinetic energy, $E_k$, using scaling relations.
The relevant basic equations are
\begin{eqnarray}
 t_r &\propto& \kappa ^{1/2} ~M_{\rm ej}^{3/4} ~E_k^{-1/4} \mbox{ ,} \mbox{ and}\\
 v &\propto& E_k^{1/2} ~M_{\rm ej}^{-1/2} \mbox{ ,}
\end{eqnarray}
where $\kappa$ is absorption coefficient for optical photons, and $v$ is a typical expansion velocity \citep{1982ApJ...253..785A}.
We apply this relation to the case of SN 2012au using the parameters derived for well-studied SN 2008D \citep{2009ApJ...692.1131T} as the scaling template.
We derive the parameters of SN 2012au as $M_{\rm ej} = 5$--$7$ M$_{\odot}$ and $E_k = (7$--$18) \times 10^{51}$ erg, where we adopt the \ion{He}{1} line velocities around maximum as the expansion velocities\footnote{When we use the velocities around $+50$ d, the parameters show only a small fractional changes as $M_{\rm ej} = 5$--$7$ M$_{\odot}$ and $E_k = (6$--$14) \times 10^{51}$ erg.}.

In Figure 5, we plot the peak $R$-band absolute magnitude $M_{R,{\rm max}}$ vs. the \ion{He}{1} $\lambda$5876 line velocity $v_{\rm He}$ around maximum (left panel) and the \Ni mass vs. the kinetic energy (right panel) in SN 2012au with those of other SNe Ib (and also SNe Ic in the right panel) for comparison. 
We see weak positive correlations both in $v_{\rm He}-M_{R,{\rm max}}$ and $E_k-M(^{56}{\rm Ni})$ except for very peculiar SN 2005bf.
To our knowledge, the former correlation has not been pointed out so far; the data point of SN Ib 2012au makes it much clearer.
Also in the right panel, the point of SN 2012au is apart from the average of SNe Ib and Ic, and rather near of the average of GRB-associated SNe.
These results may indicate that SN 2012au have observational properties similar to hypernova.

\begin{figure*}
\begin{center}
\includegraphics[scale=0.55,angle=-90]{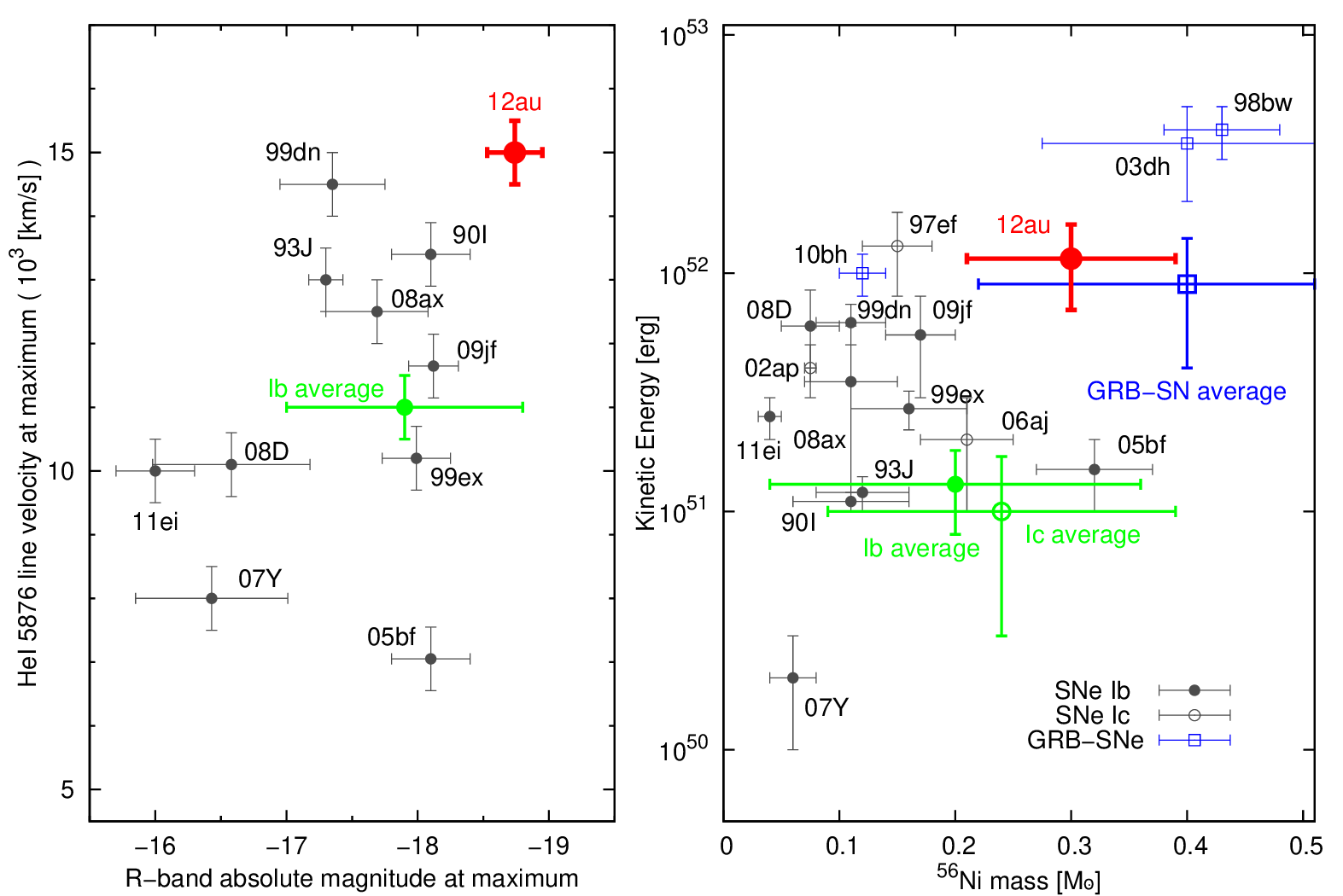}
\caption{Comparisons of physical parameters of SN 2012au with SN Ib 1990I \citep{2004A&A...426..963E}, SN IIb 1993J \citep{1995ApJ...449L..51Y}, SN Ib 1999dn \citep{2011MNRAS.411.2726B}, SN IIb 1999ex \citep{2002AJ....124.2100S}, SN Ib 2005bf \citep{2005ApJ...633L..97T,2007ApJ...666.1069M}, SN Ib 2007Y \citep{2009ApJ...696..713S}, SN Ib 2008D \citep{2009ApJ...692.1131T}, SN Ib/IIb 2008ax \citep{2011MNRAS.413.2140T}, SN Ib 2009jf \citep{2011ASInC...3..126S}, SN Ib/IIb 2011ei \citep{2013ApJ...767...71M}. The left panel shows the peak $R$-band absolute magnitude vs. the \ion{He}{1} $\lambda$5876 line velocity around maximum. The right panel shows the \Ni mass vs. the kinetic energy. We also plot the average values among SNe Ib \citep{2002ApJ...566.1005B,2011ApJ...741...97D} in both panels, and $E_k$ and $M(^{56}{\rm Ni})$ of SN Ic 1997ef \citep{2000ApJ...545..407M}, SN Ic 2002ap \citep{2002ApJ...572L..61M}, SN Ic 2006aj \citep{2006Natur.442.1018M}, the luminous SN Ic-BL `hypernova' SN 1998bw \citep{2001ApJ...550..991N}, SN 2003dh \citep{2003ApJ...599L..95M}, SN 2010bh \citep{2012ApJ...753...67B}, and the average one among SNe Ic and GRB-SNe in right panel. We can see weak positive correlations in both plots; the correlation coefficients are 0.52 and 0.58 for left and right panels respectively, except for very peculiar SN 2005bf.}
\end{center}
\end{figure*}

\section{Conclusions}
We presented early phase optical photometric and spectroscopic observations of SN 2012au and discussed that (i) SN 2012au is a very luminous SN Ib whose peak quasi-bolometric luminosity reached $\sim 6.7 \times 10^{42}$ \ergs, and that (ii) this SN shows large \ion{He} {1} velocity from maximum through $+70$ d, and (iii) there is a possible positive correlation between $v_{\rm He}$ and $M_{R,{\rm max}}$ among SNe Ib.

We derived constraints on the explosion parameters for SN 2012au as follows: $M_{\rm ej} = 5$--$7$ M$_{\odot}$ and $E_k = (7$--$18) \times 10^{51}$ erg.
While this may contain a large error, the ejecta mass points to the main-sequence mass of $M_{\rm ms} \sim 20$--$30M_{\odot}$ \citep[see][]{2009ApJ...692.1131T}.
Together with the large explosion energy and the large \Ni mass, the progenitor of SN 2012au likely had a large main sequence mass as $M_{\rm ms} > 20 M_{\odot}$, for which the outer hydrogen envelope had been stripped away but the helium layer still remained.

Although SN 2012au is spectroscopically classified as SN Ib, the bolometric luminosity is large and close to that of SN 1998bw, and the bolometric LC modeling also suggests that although SN 2012au had the helium envelope at the explosion, the structure of the ejecta in SN 2012au is similar to that of SN 1998bw.
These, together with the high velocity absorption features, suggest that SN 2012au has a character close to hypernovae.
Finally, we note that independently similar results and discussions were given by \cite{2013arXiv1304.0095M}, mostly based on later phase data than we analyzed in this Letter\footnote{At the final stage of preparing this draft, an independent work by \cite{2013arXiv1304.0095M} has appeared on arXiv.}.
This may allow us to get an important insight because no GRB-associated SNe Ib has been ever discovered.

\acknowledgements
This research has been supported in part by Optical \& Near-infrared Astronomy Inter-University Cooperation Program and by the Grant-in-Aid for Scientific Research from JSPS (23340048) 
and WPI Initiative, the MEXT of Japan.


\begin{thebibliography}{49}
\expandafter\ifx\csname natexlab\endcsname\relax\def\natexlab#1{#1}\fi

\bibitem[{{Anupama} {et~al.}(2005){Anupama}, {Sahu}, {Deng}, {Nomoto},
  {Tominaga}, {Tanaka}, {Mazzali}, \& {Prabhu}}]{2005ApJ...631L.125A}
{Anupama}, G.~C., {Sahu}, D.~K., {Deng}, J., {Nomoto}, K., {Tominaga}, N.,
  {Tanaka}, M., {Mazzali}, P.~A., \& {Prabhu}, T.~P. 2005, \apjl, 631, L125

\bibitem[{{Arnett}(1982)}]{1982ApJ...253..785A}
{Arnett}, W.~D. 1982, \apj, 253, 785

\bibitem[{{Benetti} {et~al.}(2011){Benetti}, {Turatto}, {Valenti},
  {Pastorello}, {Cappellaro}, {Botticella}, {Bufano}, {Ghinassi},
  {Harutyunyan}, {Inserra}, {Magazz{\~a}{\sup1}}, {Patat}, {Pumo}, \&
  {Taubenberger}}]{2011MNRAS.411.2726B}
{Benetti}, S., {et~al.} 2011, \mnras, 411, 2726

\bibitem[{{Branch} {et~al.}(2002){Branch}, {Benetti}, {Kasen}, {Baron},
  {Jeffery}, {Hatano}, {Stathakis}, {Filippenko}, {Matheson}, {Pastorello},
  {Altavilla}, {Cappellaro}, {Rizzi}, {Turatto}, {Li}, {Leonard}, \&
  {Shields}}]{2002ApJ...566.1005B}
{Branch}, D., {et~al.} 2002, \apj, 566, 1005

\bibitem[{{Bufano} {et~al.}(2012){Bufano}, {Pian}, {Sollerman}, {Benetti},
  {Pignata}, {Valenti}, {Covino}, {D'Avanzo}, {Malesani}, {Cappellaro}, {Della
  Valle}, {Fynbo}, {Hjorth}, {Mazzali}, {Reichart}, {Starling}, {Turatto},
  {Vergani}, {Wiersema}, {Amati}, {Bersier}, {Campana}, {Cano},
  {Castro-Tirado}, {Chincarini}, {D'Elia}, {de Ugarte Postigo}, {Deng},
  {Ferrero}, {Filippenko}, {Goldoni}, {Gorosabel}, {Greiner}, {Hammer},
  {Jakobsson}, {Kaper}, {Kawabata}, {Klose}, {Levan}, {Maeda}, {Masetti},
  {Milvang-Jensen}, {Mirabel}, {M{\o}ller}, {Nomoto}, {Palazzi}, {Piranomonte},
  {Salvaterra}, {Stratta}, {Tagliaferri}, {Tanaka}, {Tanvir}, \&
  {Wijers}}]{2012ApJ...753...67B}
{Bufano}, F., {et~al.} 2012, \apj, 753, 67

\bibitem[{{Clocchiatti} \& {Wheeler}(1997)}]{1997ApJ...491..375C}
{Clocchiatti}, A., \& {Wheeler}, J.~C. 1997, \apj, 491, 375

\bibitem[{{Crockett} {et~al.}(2008){Crockett}, {Eldridge}, {Smartt},
  {Pastorello}, {Gal-Yam}, {Fox}, {Leonard}, {Kasliwal}, {Mattila}, {Maund},
  {Stephens}, \& {Danziger}}]{2008MNRAS.391L...5C}
{Crockett}, R.~M., {et~al.} 2008, \mnras, 391, L5

\bibitem[{{Deng} {et~al.}(2000){Deng}, {Qiu}, {Hu}, {Hatano}, \&
  {Branch}}]{2000ApJ...540..452D}
{Deng}, J.~S., {Qiu}, Y.~L., {Hu}, J.~Y., {Hatano}, K., \& {Branch}, D. 2000,
  \apj, 540, 452

\bibitem[{{Drout} {et~al.}(2011){Drout}, {Soderberg}, {Gal-Yam}, {Cenko},
  {Fox}, {Leonard}, {Sand}, {Moon}, {Arcavi}, \& {Green}}]{2011ApJ...741...97D}
{Drout}, M.~R., {et~al.} 2011, \apj, 741, 97

\bibitem[{{Elmhamdi} {et~al.}(2004){Elmhamdi}, {Danziger}, {Cappellaro}, {Della
  Valle}, {Gouiffes}, {Phillips}, \& {Turatto}}]{2004A&A...426..963E}
{Elmhamdi}, A., {Danziger}, I.~J., {Cappellaro}, E., {Della Valle}, M.,
  {Gouiffes}, C., {Phillips}, M.~M., \& {Turatto}, M. 2004, \aap, 426, 963

\bibitem[{{Filippenko}(1997)}]{1997ARA&A..35..309F}
{Filippenko}, A.~V. 1997, \araa, 35, 309

\bibitem[{{Hamuy} {et~al.}(2002){Hamuy}, {Maza}, {Pinto}, {Phillips},
  {Suntzeff}, {Blum}, {Olsen}, {Pinfield}, {Ivanov}, {Augusteijn}, {Brillant},
  {Chadid}, {Cuby}, {Doublier}, {Hainaut}, {Le Floc'h}, {Lidman},
  {Petr-Gotzens}, {Pompei}, \& {Vanzi}}]{2002AJ....124..417H}
{Hamuy}, M., {et~al.} 2002, \aj, 124, 417

\bibitem[{{Howell} {et~al.}(2006){Howell}, {Sullivan}, {Nugent}, {Ellis},
  {Conley}, {Le Borgne}, {Carlberg}, {Guy}, {Balam}, {Basa}, {Fouchez}, {Hook},
  {Hsiao}, {Neill}, {Pain}, {Perrett}, \& {Pritchet}}]{2006Natur.443..308H}
{Howell}, D.~A., {et~al.} 2006, \nat, 443, 308

\bibitem[{{Howerton} {et~al.}(2012){Howerton}, {Drake}, {Djorgovski},
  {Mahabal}, {Graham}, {Williams}, {Roy}, {Mohan}, {Prieto}, {Catelan},
  {Beshore}, {Larson}, {Christensen}, {Elenin}, {Molotov}, {Koff}, {Silverman},
  {Cenko}, {Miller}, {Nugent}, \& {Filippenko}}]{2012CBET.3052....1H}
{Howerton}, S., {et~al.} 2012, Central Bureau Electronic Telegrams, 3052, 1

\bibitem[{{Kawabata} {et~al.}(2010){Kawabata}, {Maeda}, {Nomoto},
  {Taubenberger}, {Tanaka}, {Deng}, {Pian}, {Hattori}, \&
  {Itagaki}}]{2010Natur.465..326K}
{Kawabata}, K.~S., {et~al.} 2010, \nat, 465, 326

\bibitem[{{Kawabata} {et~al.}(2008){Kawabata}, {Nagae}, {Chiyonobu}, {Tanaka},
  {Nakaya}, {Suzuki}, {Kamata}, {Miyazaki}, {Hiragi}, {Miyamoto}, {Yamanaka},
  {Arai}, {Yamashita}, {Uemura}, {Ohsugi}, {Isogai}, {Ishitobi}, \&
  {Sato}}]{2008SPIE.7014E.151K}
{Kawabata}, K.~S., {et~al.} 2008, in Society of Photo-Optical Instrumentation
  Engineers (SPIE) Conference Series, Vol. 7014, Society of Photo-Optical
  Instrumentation Engineers (SPIE) Conference Series

\bibitem[{{Li} {et~al.}(2011){Li}, {Leaman}, {Chornock}, {Filippenko},
  {Poznanski}, {Ganeshalingam}, {Wang}, {Modjaz}, {Jha}, {Foley}, \&
  {Smith}}]{2011MNRAS.412.1441L}
{Li}, W., {et~al.} 2011, \mnras, 412, 1441

\bibitem[{{Maeda} {et~al.}(2003){Maeda}, {Mazzali}, {Deng}, {Nomoto}, {Yoshii},
  {Tomita}, \& {Kobayashi}}]{2003ApJ...593..931M}
{Maeda}, K., {Mazzali}, P.~A., {Deng}, J., {Nomoto}, K., {Yoshii}, Y.,
  {Tomita}, H., \& {Kobayashi}, Y. 2003, \apj, 593, 931

\bibitem[{{Maeda} {et~al.}(2007){Maeda}, {Tanaka}, {Nomoto}, {Tominaga},
  {Kawabata}, {Mazzali}, {Umeda}, {Suzuki}, \& {Hattori}}]{2007ApJ...666.1069M}
{Maeda}, K., {et~al.} 2007, \apj, 666, 1069

\bibitem[{{Matheson} {et~al.}(2001){Matheson}, {Filippenko}, {Li}, {Leonard},
  \& {Shields}}]{2001AJ....121.1648M}
{Matheson}, T., {Filippenko}, A.~V., {Li}, W., {Leonard}, D.~C., \& {Shields},
  J.~C. 2001, \aj, 121, 1648

\bibitem[{{Maund} \& {Smartt}(2005)}]{2005MNRAS.360..288M}
{Maund}, J.~R., \& {Smartt}, S.~J. 2005, \mnras, 360, 288

\bibitem[{{Mazzali} {et~al.}(2002){Mazzali}, {Deng}, {Maeda}, {Nomoto},
  {Umeda}, {Hatano}, {Iwamoto}, {Yoshii}, {Kobayashi}, {Minezaki}, {Doi},
  {Enya}, {Tomita}, {Smartt}, {Kinugasa}, {Kawakita}, {Ayani}, {Kawabata},
  {Yamaoka}, {Qiu}, {Motohara}, {Gerardy}, {Fesen}, {Kawabata}, {Iye},
  {Kashikawa}, {Kosugi}, {Ohyama}, {Takada-Hidai}, {Zhao}, {Chornock},
  {Filippenko}, {Benetti}, \& {Turatto}}]{2002ApJ...572L..61M}
{Mazzali}, P.~A., {et~al.} 2002, \apjl, 572, L61

\bibitem[{{Mazzali} {et~al.}(2006){Mazzali}, {Deng}, {Nomoto}, {Sauer}, {Pian},
  {Tominaga}, {Tanaka}, {Maeda}, \& {Filippenko}}]{2006Natur.442.1018M}
---. 2006, \nat, 442, 1018

\bibitem[{{Mazzali} {et~al.}(2003){Mazzali}, {Deng}, {Tominaga}, {Maeda},
  {Nomoto}, {Matheson}, {Kawabata}, {Stanek}, \&
  {Garnavich}}]{2003ApJ...599L..95M}
---. 2003, \apjl, 599, L95

\bibitem[{{Mazzali} {et~al.}(2000){Mazzali}, {Iwamoto}, \&
  {Nomoto}}]{2000ApJ...545..407M}
{Mazzali}, P.~A., {Iwamoto}, K., \& {Nomoto}, K. 2000, \apj, 545, 407

\bibitem[{{Mazzali} {et~al.}(2008){Mazzali}, {Valenti}, {Della Valle},
  {Chincarini}, {Sauer}, {Benetti}, {Pian}, {Piran}, {D'Elia}, {Elias-Rosa},
  {Margutti}, {Pasotti}, {Antonelli}, {Bufano}, {Campana}, {Cappellaro},
  {Covino}, {D'Avanzo}, {Fiore}, {Fugazza}, {Gilmozzi}, {Hunter}, {Maguire},
  {Maiorano}, {Marziani}, {Masetti}, {Mirabel}, {Navasardyan}, {Nomoto},
  {Palazzi}, {Pastorello}, {Panagia}, {Pellizza}, {Sari}, {Smartt},
  {Tagliaferri}, {Tanaka}, {Taubenberger}, {Tominaga}, {Trundle}, \&
  {Turatto}}]{2008Sci...321.1185M}
{Mazzali}, P.~A., {et~al.} 2008, Science, 321, 1185

\bibitem[{{Milisavljevic} {et~al.}(2013{\natexlab{a}}){Milisavljevic},
  {Margutti}, {Soderberg}, {Pignata}, {Chomiuk}, {Fesen}, {Bufano}, {Sanders},
  {Parrent}, {Parker}, {Mazzali}, {Pian}, {Pickering}, {Buckley}, {Crawford},
  {Gulbis}, {Hettlage}, {Hooper}, {Nordsieck}, {O'Donoghue}, {Husser},
  {Potter}, {Kniazev}, {Kotze}, {Romero-Colmenero}, {Vaisanen}, {Wolf},
  {Bietenholz}, {Bartel}, {Fransson}, {Walker}, {Brunthaler}, {Chakraborti},
  {Levesque}, {MacFadyen}, {Drescher}, {Bock}, {Marples}, {Anderson},
  {Benetti}, {Reichart}, \& {Ivarsen}}]{2013ApJ...767...71M}
{Milisavljevic}, D., {et~al.} 2013{\natexlab{a}}, \apj, 767, 71

\bibitem[{{Milisavljevic} {et~al.}(2013{\natexlab{b}}){Milisavljevic},
  {Soderberg}, {Margutti}, {Drout}, {Marion}, {Sanders}, {Hsiao}, {Lunnan},
  {Chornock}, {Fesen}, {Parrent}, {Levesque}, {Berger}, {Foley}, {Challis},
  {Kirshner}, {Dittmann}, {Bieryla}, {Kamble}, {Chakroborti}, {De Rosa},
  {Fausnaugh}, {Hainline}, {Chen}, {Hickox}, {Morrell}, {Phillips}, \&
  {Stritzinger}}]{2013arXiv1304.0095M}
---. 2013{\natexlab{b}}, ArXiv e-prints

\bibitem[{{Modjaz} {et~al.}(2009){Modjaz}, {Li}, {Butler}, {Chornock},
  {Perley}, {Blondin}, {Bloom}, {Filippenko}, {Kirshner}, {Kocevski},
  {Poznanski}, {Hicken}, {Foley}, {Stringfellow}, {Berlind}, {Barrado y
  Navascues}, {Blake}, {Bouy}, {Brown}, {Challis}, {Chen}, {de Vries},
  {Dufour}, {Falco}, {Friedman}, {Ganeshalingam}, {Garnavich}, {Holden},
  {Illingworth}, {Lee}, {Liebert}, {Marion}, {Olivier}, {Prochaska},
  {Silverman}, {Smith}, {Starr}, {Steele}, {Stockton}, {Williams}, \&
  {Wood-Vasey}}]{2009ApJ...702..226M}
{Modjaz}, M., {et~al.} 2009, \apj, 702, 226

\bibitem[{{Nakamura} {et~al.}(2001){Nakamura}, {Mazzali}, {Nomoto}, \&
  {Iwamoto}}]{2001ApJ...550..991N}
{Nakamura}, T., {Mazzali}, P.~A., {Nomoto}, K., \& {Iwamoto}, K. 2001, \apj,
  550, 991

\bibitem[{{Patat} {et~al.}(2001){Patat}, {Cappellaro}, {Danziger}, {Mazzali},
  {Sollerman}, {Augusteijn}, {Brewer}, {Doublier}, {Gonzalez}, {Hainaut},
  {Lidman}, {Leibundgut}, {Nomoto}, {Nakamura}, {Spyromilio}, {Rizzi},
  {Turatto}, {Walsh}, {Galama}, {van Paradijs}, {Kouveliotou}, {Vreeswijk},
  {Frontera}, {Masetti}, {Palazzi}, \& {Pian}}]{2001ApJ...555..900P}
{Patat}, F., {et~al.} 2001, \apj, 555, 900

\bibitem[{{Perets} {et~al.}(2010){Perets}, {Gal-Yam}, {Mazzali}, {Arnett},
  {Kagan}, {Filippenko}, {Li}, {Arcavi}, {Cenko}, {Fox}, {Leonard}, {Moon},
  {Sand}, {Soderberg}, {Anderson}, {James}, {Foley}, {Ganeshalingam}, {Ofek},
  {Bildsten}, {Nelemans}, {Shen}, {Weinberg}, {Metzger}, {Piro}, {Quataert},
  {Kiewe}, \& {Poznanski}}]{2010Natur.465..322P}
{Perets}, H.~B., {et~al.} 2010, \nat, 465, 322

\bibitem[{{Poznanski} {et~al.}(2012){Poznanski}, {Prochaska}, \&
  {Bloom}}]{2012MNRAS.426.1465P}
{Poznanski}, D., {Prochaska}, J.~X., \& {Bloom}, J.~S. 2012, \mnras, 426, 1465

\bibitem[{{Richmond} {et~al.}(1996){Richmond}, {Treffers}, {Filippenko}, \&
  {Paik}}]{1996AJ....112..732R}
{Richmond}, M.~W., {Treffers}, R.~R., {Filippenko}, A.~V., \& {Paik}, Y. 1996,
  \aj, 112, 732

\bibitem[{{Sahu} {et~al.}(2011){Sahu}, {Gurugubelli}, {Anupama}, \&
  {Nomoto}}]{2011ASInC...3..126S}
{Sahu}, D.~K., {Gurugubelli}, U.~K., {Anupama}, G.~C., \& {Nomoto}, K. 2011, in
  Astronomical Society of India Conference Series, Vol.~3, Astronomical Society
  of India Conference Series, 126

\bibitem[{{Schlegel} {et~al.}(1998){Schlegel}, {Finkbeiner}, \&
  {Davis}}]{1998ApJ...500..525S}
{Schlegel}, D.~J., {Finkbeiner}, D.~P., \& {Davis}, M. 1998, \apj, 500, 525

\bibitem[{{Silverman} {et~al.}(2012){Silverman}, {Cenko}, {Miller}, {Nugent},
  \& {Filippenko}}]{2012CBET.3052....2S}
{Silverman}, J.~M., {Cenko}, S.~B., {Miller}, A.~A., {Nugent}, P.~E., \&
  {Filippenko}, A.~V. 2012, Central Bureau Electronic Telegrams, 3052, 2

\bibitem[{{Soderberg} {et~al.}(2012){Soderberg}, {Dittmann}, {Claus}, {Esty},
  {Harrison}, {Kurcz}, {Michaels}, {Oprescu}, {Powell}, {Speagle},
  {Suleynanzade}, {Tu}, \& {Wolansky}}]{2012ATel.3968....1S}
{Soderberg}, A., {et~al.} 2012, The Astronomer's Telegram, 3968, 1

\bibitem[{{Stritzinger} {et~al.}(2002){Stritzinger}, {Hamuy}, {Suntzeff},
  {Smith}, {Phillips}, {Maza}, {Strolger}, {Antezana}, {Gonz{\'a}lez},
  {Wischnjewsky}, {Candia}, {Espinoza}, {Gonz{\'a}lez}, {Stubbs}, {Becker},
  {Rubenstein}, \& {Galaz}}]{2002AJ....124.2100S}
{Stritzinger}, M., {et~al.} 2002, \aj, 124, 2100

\bibitem[{{Stritzinger} {et~al.}(2009){Stritzinger}, {Mazzali}, {Phillips},
  {Immler}, {Soderberg}, {Sollerman}, {Boldt}, {Braithwaite}, {Brown}, {Burns},
  {Contreras}, {Covarrubias}, {Folatelli}, {Freedman}, {Gonz{\'a}lez}, {Hamuy},
  {Krzeminski}, {Madore}, {Milne}, {Morrell}, {Persson}, {Roth}, {Smith}, \&
  {Suntzeff}}]{2009ApJ...696..713S}
---. 2009, \apj, 696, 713

\bibitem[{{Stritzinger} {et~al.}(2006){Stritzinger}, {Mazzali}, {Sollerman}, \&
  {Benetti}}]{2006A&A...460..793S}
{Stritzinger}, M., {Mazzali}, P.~A., {Sollerman}, J., \& {Benetti}, S. 2006,
  \aap, 460, 793

\bibitem[{{Tanaka} {et~al.}(2009){Tanaka}, {Tominaga}, {Nomoto}, {Valenti},
  {Sahu}, {Minezaki}, {Yoshii}, {Yoshida}, {Anupama}, {Benetti}, {Chincarini},
  {Della Valle}, {Mazzali}, \& {Pian}}]{2009ApJ...692.1131T}
{Tanaka}, M., {et~al.} 2009, \apj, 692, 1131

\bibitem[{{Taubenberger} {et~al.}(2011){Taubenberger}, {Navasardyan}, {Maurer},
  {Zampieri}, {Chugai}, {Benetti}, {Agnoletto}, {Bufano}, {Elias-Rosa},
  {Turatto}, {Patat}, {Cappellaro}, {Mazzali}, {Iijima}, {Valenti},
  {Harutyunyan}, {Claudi}, \& {Dolci}}]{2011MNRAS.413.2140T}
{Taubenberger}, S., {et~al.} 2011, \mnras, 413, 2140

\bibitem[{{Theureau} {et~al.}(2007){Theureau}, {Hanski}, {Coudreau}, {Hallet},
  \& {Martin}}]{2007A&A...465...71T}
{Theureau}, G., {Hanski}, M.~O., {Coudreau}, N., {Hallet}, N., \& {Martin},
  J.-M. 2007, \aap, 465, 71

\bibitem[{{Tominaga} {et~al.}(2005){Tominaga}, {Tanaka}, {Nomoto}, {Mazzali},
  {Deng}, {Maeda}, {Umeda}, {Modjaz}, {Hicken}, {Challis}, {Kirshner},
  {Wood-Vasey}, {Blake}, {Bloom}, {Skrutskie}, {Szentgyorgyi}, {Falco},
  {Inada}, {Minezaki}, {Yoshii}, {Kawabata}, {Iye}, {Anupama}, {Sahu}, \&
  {Prabhu}}]{2005ApJ...633L..97T}
{Tominaga}, N., {et~al.} 2005, \apjl, 633, L97

\bibitem[{{Tomita} {et~al.}(2006){Tomita}, {Deng}, {Maeda}, {Yoshii}, {Nomoto},
  {Mazzali}, {Suzuki}, {Kobayashi}, {Minezaki}, {Aoki}, {Enya}, \&
  {Suganuma}}]{2006ApJ...644..400T}
{Tomita}, H., {et~al.} 2006, \apj, 644, 400

\bibitem[{{Tsvetkov} {et~al.}(2009){Tsvetkov}, {Volkov}, {Baklanov},
  {Blinnikov}, \& {Tuchin}}]{2009PZ.....29....2T}
{Tsvetkov}, D.~Y., {Volkov}, I.~M., {Baklanov}, P., {Blinnikov}, S., \&
  {Tuchin}, O. 2009, Peremennye Zvezdy, 29, 2

\bibitem[{{Tully}(1988)}]{1988Sci...242..310T}
{Tully}, R.~B. 1988, Science, 242, 310

\bibitem[{{Young} {et~al.}(1995){Young}, {Baron}, \&
  {Branch}}]{1995ApJ...449L..51Y}
{Young}, T.~R., {Baron}, E., \& {Branch}, D. 1995, \apjl, 449, L51

\end{thebibliography}

\end{document}